\documentstyle[twocolumn]{jpsj}
\newcommand{\nolig}{\textcompwordmark}
\newcommand{\rc}{{\rm c}}
\newcommand{\rf}{{\rm f}}

\newcommand{\ep}{\epsilon}
\newcommand{\Ufc}{U_{\rf\rc}}

\def\runtitle
{
Huge Enhancement of Impurity Scattering due to Critical Valence 
Fluctuations 
}
\def\runauthor
{
Kazumasa {\sc Miyake} and Hideaki {\sc Maebashi}
}

\title{
Huge Enhancement of Impurity Scattering due to Critical Valence Fluctuations 
in a Ce-Based Heavy Electron System 
}

\author{
Kazumasa {\sc Miyake} and Hideaki {\sc Maebashi}
}

\inst
{Division of Materials Physics, Department of Physical Science, 
Graduate School of Engineering Science, 
Osaka University, Toyonaka, Osaka 560-8531, Japan
}
\recdate
{
\today
}

\abst
{On the basis of the Ward-Pitaevskii identity, the residual resistivity 
$\rho_{0}$ is shown to exhibit huge enhancement around the quantum critical 
point of valence transition in Ce-based heavy electron systems.  
This explains a sharp peak of $\rho_{0}$ 
observed in CeCu$_2$Ge$_2$ under the pressure at $P\sim$16GPa where 
the superconducting trasition temperature also exhibit the sharp peak.  
}

\kword
{enhanced impurity scattering, critical valence fluctuations, 
CeCu$_2$Ge$_2$, CeCu$_2$Si$_2$, 
}

\input epsf.tex
\begin{document}
\sloppy
\maketitle
%\newpage
Recently, renormalization effect of impurity potential by quantum 
critical fluctuations has begun to attract much 
attention\cite{Kotliar,varma,MNO,MN,Jaccard,Thessieu,MMV}, while the effect 
of impurities on the universality class of critical fluctuations was 
clarified quite long ago\cite{Fulde}, and that on the temperature dependence 
of the resistivity at quantum criticality has been discussed 
recently\cite{rosch}.  
A theoretical guideline for discussing exactly such an effect 
has already been put forth more than three decades ago by Betbder-Matibet 
and Nozi\`eres in the framework of the Fermi liquid theory\cite{Matibet}.  
Indeed, they 
showed on the basis of the Ward identy that the impurity potenital, 
in one-component Fermi liquid, is 
renormalized by many-body effect in the forward scattering limit as 
\begin{equation}
\label{eq:1}
{\tilde u}({\vec k}\to 0)={1\over z(1+F_{0}^{\rm s})}
u({\vec k}\to 0),
\end{equation}
where $u({\vec k})$ is the bare non-magnetic impurity potential and 
${\tilde u}({\vec k})$ are the renormalized one, $z$ is the 
renormalization amplitude including {\it all} the manybody 
effects, and $F_{0}^{\rm s}$ the Landau parameter relevant to the 
correction of the charge susceptibility.  For the 
potential of magnetic impurity, the relation similar to (\ref{eq:1}) holds 
with the Fermi liquid parameter $F_{0}^{\rm s}$ being replaced by 
$F_{0}^{\rm a}$ which gives the Fermi liquid correction of the spin 
susceptibility.  

Heavy electron compound CeCu$_2$Ge$_2$ at ambient pressure changes 
drastically its electronic state at the pressure $P\simeq17$GPa where 
the coefficient $A$ of $T^{2}$-term of the resistivity $\rho(T)$ decreases 
by about three orders of magnitude and the universal ratio 
$A/\gamma^{2}$, $\gamma$ being the Sommerfeld constant, changes from the 
value of heavy electorns to that of conventional $d$-band metals decreasing 
by 25 times\cite{Jaccard}.  
This suggests that the rapid valence change of Ce ion occurs at around that 
pressure.  Indeed, the rapid volume change maintaining the crystal symmetry 
was observed at around the ``critical" pressure by SOR-X ray diffraction 
implying that the rapid valence change occurs there\cite{onodera}.  
It was also observed that the residual resistivity  $\rho_{0}$ 
exhibits sharp peak at the same pressure\cite{Jaccard}. 
Similar behavior has been observed 
in CeRhIn$_5$ recently discovered pressure induce 
superconductor\cite{Muramatsu}.  Recently, it was shown that 
$\rho_{0}$ can be enhanced through the renormalization of impurity 
potential by exchanging the critical valence 
fluctuations\cite{MNO,MN,MM}.  However, 
the argument is based on the perturbational treatment with the use of 
the phenomenological form of valence fluctuation propagator.   A purpose of 
this {\sc Letter} is to extend that treatment so as 
to take into account the full effect of vertex corrections to the 
impurity potenital on the basis of the Ward-Pitaevskii identity in 
two-component Fermi liquid.  
Namely, we derive a technically exact formula for 
the renormaliztion of impurity potential for the forward scattering 
associated with a rapid critical valence change in heavy electrons 
form the Kondo regime to the valence fluctuation one.  

We start with an extended periodic Anderson model (PAM),
\begin{eqnarray}
& &H=\sum_{p \sigma}\xi_k c_{p \sigma}^{\dagger}c_{p \sigma}^{}
 +\ep_\rf \sum_{p \sigma}f_{p \sigma}^{\dagger}f_{p \sigma}^{} 
 +U_{\rf\nolig\rf}\sum_i n_{i \uparrow}^\rf n_{i \downarrow}^\rf 
\nonumber \\
& &\qquad
 +\sum_{p \sigma}(V_{p}c_{p \sigma}^{\dagger}f_{p \sigma}^{}+{\rm h.c.})
 +\Ufc\sum_{i \sigma \sigma'}n_{i \sigma}^{\rm f}n_{i \sigma'}^\rc,
 \label{eq:PAM}
\end{eqnarray}
where the conventional notations for PAM are used except for $U_{\rm fc}$, 
the f-c Coulomb repulsion.  It has recently been shown that 
the effect of $\Ufc$ is important for the rapid valence change to 
occur as the f-level $\ep_\rf$ is increased approaching the Fermi level 
by the effect of pressure\cite{Onishi1,Onishi2}.  
The one-particle Green function for a given spin direction in 
this system is given formally as 
\begin{eqnarray}
& & \left[G^{-1}_{ij}({\vec p},\varepsilon)\right]
=  
\left[
\begin{array}{cc}
G_{\rf\rf}({\vec p},\varepsilon) & G_{\rf\rc}({\vec p},\varepsilon) \\
G_{\rc\rf}({\vec p},\varepsilon) & G_{\rc\rc}({\vec p},\varepsilon) 
\end{array}
\right]^{-1} 
\label{eq:Gf1}
\\
& & =
\left[
\begin{array}{cc}
\varepsilon-\epsilon_{\rf}+\mu-\Sigma_{\rf\rf}({\vec p},\varepsilon) & 
-V_{p}-\Sigma_{\rf\rc}({\vec p},\varepsilon) \\
-V_{p}^*-\Sigma_{\rc\rf}({\vec p},\varepsilon) & 
\varepsilon-\xi_{\vec p}-\Sigma_{\rc\rc}({\vec p},\varepsilon)
\end{array}
\right],
\label{eq:Gf2}
\end{eqnarray}
where $\Sigma_{\rf\rf}$ is the selfenergy of f-electron and
there also exist $\Sigma_{\rf\rc}$ and $\Sigma_{\rc\rc}$
as the many-body effect due to $U_{\rm ff}$ and $U_{\rm fc}$. 
It is noted that $G^{-1}_{ij}$ means $ij$-element of the inverse matrix 
of the Green function, e.g. $G^{-1}_{11} \neq G^{-1}_{\rf\rf}$ while 
$G_{11} = G_{\rf\rf}$. 

For coupling to nonmagnetic impurities, the scattering matrix 
for this system is given generally as 
$u_{ij}({\vec k}) = \sum_{a=1}^4 u_a({\vec k}) \lambda^a_{ij}$ where 
$u_1$ and $u_2$ are the variations of potential
on f-electrons and c-electrons, respectively,
while $u_3$ and $u_4$ represent the strength of the f-c mixing scattering. 
Here $\lambda^a_{ij}$ is the bare vertex for coupling to impurities and 
is given by 
$[\lambda^1_{ij}] = (1+\tau^z)/2$, 
$[\lambda^2_{ij}] = (1-\tau^z)/2$, 
$[\lambda^3_{ij}] = \tau^x/\sqrt{2}$ 
and 
$[\lambda^4_{ij}] = \tau^y/\sqrt{2}$ 
with ${\vec \tau}$ being the Pauli matrices in the f-c space.  

In the following, a Greek index, e.g. $\alpha$, represents 
the dependence on both the component $i = 1,2$ 
and the spin $\sigma = \uparrow, \downarrow$, and 
the summation is assumed to be taken for repeated indices. 
Then, the one-particle Green function and the scattering matrix 
mean the tensor product multiplied by 
the unit matrix with respect to the spin variables.  

\begin{figure}[htbp]
\begin{center}
\epsfxsize=7.5cm \epsfbox{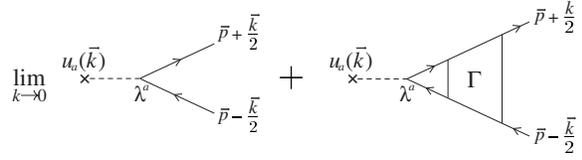}
\end{center}
\caption{
Feynman diagram for the exact vertex correction of impurity 
potential $u({\vec k})$. 
}
\label{fig:1}
\end{figure}

The Ward-Pitaevskii identity relevant to the present problem 
is given by considering the linear response of 
$G_{\gamma\alpha}$ caused by the shift of the parameter 
denoted by $\ep_{a}$ 
with the chemical potential $\mu$ being fixed. 
Here $\ep_{a}$ is 
the f-level $\ep_\rf$, the center of the conduction band $\ep_{\rc}$ 
or f-c mixing $V$, i.e.,
$^t[\ep_a]=(\ep_1, \ep_2, \ep_3, \ep_4) =
(\ep_\rf, \ep_{\rc}, \sqrt{2}V', \sqrt{2}V'')$ 
with $V'$ and $V''$ being the real and imaginary part of $V$, 
respectively. 
One can show, by anlyzing the structure of 
perturbation series of the selfenergy, 
that the following identity holds:
\begin{eqnarray}
& &
-\bigg( 
\frac{\partial G_{\gamma\alpha}^{-1}(p)
}{\partial\epsilon_{a}}
\bigg)_{\mu}
=\lambda^{a}_{\gamma\alpha}+
\bigg({\partial\Sigma_{\gamma\alpha}(p)
\over\partial\epsilon_{a}}
\bigg)_{\mu}
\label{eq:Ward1}\\
& &=\lambda^{a}_{\gamma\alpha}
-{\rm i}
\int{{\rm d}^4q\over(2\pi)^{4}}
\Gamma^{k}_{\gamma \delta, \alpha \beta}(p,q)
\{G_{\beta \zeta}(q)G_{\kappa \delta}(q)\}_k
\lambda^{a}_{\zeta\kappa}
\quad
\label{eq:Ward2}
\end{eqnarray}
where $\Gamma^{k}$ is the so-called $k$-limit of the full vertex function, 
and $\{GG\}_{k}$ is the same limit of particle-hole Green function 
pair with the four-vector abbreviations $p = ({\vec p},\varepsilon)$, etc. 
The process of renormalization of impurity potential is 
represented by the Feynman diagram as shown in Fig.~\ref{fig:1}.  
In the limit of 
forward scattering, i.e., $k\to 0$, the renormalized potential 
${\tilde u}_{a}({\vec k})$ and 
the bare one $u_{a}({\vec k})$ are in the relation 
\begin{eqnarray}
& &\lim_{k\to 0}{\tilde u}_{a}({\vec k})
=\lim_{k\to 0}\frac{1}{2}\lambda^{a}_{\alpha\gamma}
u_{b}({\vec k})
\nonumber \\
& &\times
\bigg[
\lambda^{b}_{\gamma\alpha}-{\rm i}
\int{{\rm d}^4q\over(2\pi)^{4}}
\Gamma^{k}_{\gamma \delta, \alpha \beta}(p,q)
\{G_{\beta \zeta}(q)G_{\kappa \delta}(q)\}_k
\lambda^{b}_{\zeta\kappa}\bigg].
\quad\,\,
\label{eq:repotential}
\end{eqnarray}
The reason why the $k$-limit vertex appears in (\ref{eq:repotential}) is 
that the impurity scattering is elastic maintaining the energy transfer 
$\omega=0$.  Therefore, by the relation (\ref{eq:Ward2}), the impurity 
potential is renormalized by 
$-(\partial G_{\gamma\alpha}^{-1}({\vec p},\varepsilon)/
\partial\epsilon_{a})_{\mu}$ in the limit $k\to 0$. 

In heavy electrons, 
the important effect of impurity on the quasiparticles 
arises from the variations of potential on f-electrons, 
because the quasiparticles consist mainly of f-electrons. 
Of these effects, those by displacement of non-f elements from the 
regular alignment around Ce ions is subject to remarkable renormalization 
by the critical valence fluctuations, because the impurity potential 
due to such effects is off the unitarity limit and has a space to be 
renormalized furthermore.  On the other hand, the defect of Ce ions gives 
rise to the unitarity scattering from the beginning in heavy electron state 
so that its potential is subject only to a gradual renormalization with a 
weak anomaly around a possible transition point from the Kondo regime to 
VF one.  
It is noted that even if we consider only the effect of the shift 
of the f-level, i.e. in the case of $u_2=u_3=u_4=0$, 
the many-body effect due to $U_{\rf\rc}$ gives rise to 
the effective f-c and c-c scattering as can be seen in (\ref{eq:repotential}).

In the present system described by the Hamiltonian (\ref{eq:PAM}), 
the total number density of f-electrons and conduction electrons 
is conserved.
This leads to the following identity:
\begin{eqnarray}
& &
\frac{\partial G^{-1}_{\gamma\alpha}(p)
}{\partial\varepsilon}
=\delta_{\gamma\alpha}-{\partial\Sigma_{\gamma\alpha}(p)
\over\partial\varepsilon}
\\
& &=\delta_{\gamma\alpha}-{\rm i}
\int{{\rm d}^4q\over(2\pi)^{4}}
\Gamma^{\omega}_{\gamma \delta, \alpha \beta}(p,q)
\{G_{\beta \zeta}(q)G_{\zeta \delta}(q)\}_\omega
\quad
\label{eq:Ward3}
\end{eqnarray}
where $\Gamma^{\omega}$ is the so-called 
$\omega$-limit of the full vertex function 
and $\{GG\}_{\omega}$ 
is the same limit of particle-hole Green function pair. 
The relation between $\Gamma^{k}$ and $\Gamma^{\omega}$ is given by
\begin{eqnarray}
& &\Gamma^{k}_{\gamma \delta, \alpha \beta}(p,p^{\prime}) 
= \Gamma^{\omega}_{\gamma \delta, \alpha \beta}(p,p^{\prime})
-{\rm i}\int{{\rm d}^4q\over(2\pi)^{4}}
\Gamma_{\gamma \kappa, \alpha \zeta}^{\omega}(p,q)
\nonumber \\
& & \times
\left[ 
\{G_{\zeta\xi}(q)
G_{\eta\kappa}(q)\}_{k}
-\{G_{\zeta\xi}(q)
G_{\eta\kappa}(q)\}_{\omega}
\right]
\Gamma^{k}_{\xi \delta, \eta \beta}(q,p^{\prime}). 
\quad\quad 
\label{eq:BS}
\end{eqnarray}
By using formulas (\ref{eq:Ward3}) and (\ref{eq:BS}), 
we can evaluate the left-hand side of (\ref{eq:Ward1})
as shown below. 

First, we write the total number density $n$ 
in terms of the Green function as
\begin{equation}
n = - {\rm i}\int{{\rm d}^4p\over(2\pi)^{4}}G_{\alpha\alpha}(p).
\label{eq:number} 
\end{equation}
Differentiation of (\ref{eq:number}) with respect to $\epsilon_{a}$ 
with $\mu$ being fixed gives
\begin{equation}
\bigg({\partial n \over\partial\epsilon_{a}}\bigg)_{\mu}
=  {\rm i}\int{{\rm d}^4p\over(2\pi)^{4}}
\{G_{\alpha\beta}(p)
G_{\beta\gamma}(p)\}_k
\bigg(\frac{\partial G_{\gamma\alpha}^{-1}(p)
}{\partial\epsilon_{a}}
\bigg)_{\mu}.
\label{eq:partn}
\end{equation}
By (\ref{eq:Ward2}) and (\ref{eq:BS}), 
we find the right hand side of (\ref{eq:partn}) is
\begin{eqnarray}
& &
{\rm i}\int{{\rm d}^4p\over(2\pi)^{4}}
\bigg[
\delta_{\kappa\zeta}
-{\rm i}\int{{\rm d}^4q\over(2\pi)^{4}}
\{G_{\alpha\beta}(q)
G_{\beta\gamma}(q)\}_{\omega}
\Gamma^{\omega}_{\gamma \kappa, \alpha \zeta}(q, p)
\bigg]
\nonumber
\\
& & \times
\left[ 
\{G_{\zeta\xi}(p)
G_{\eta\kappa}(p)\}_{k}
-\{G_{\zeta\xi}(p)
G_{\eta\kappa}(p)\}_{\omega}
\right]
\bigg(\frac{\partial G_{\xi\eta}^{-1}(p)
}{\partial\epsilon_{a}}
\bigg)_{\mu}
\nonumber \\
& &
-{\rm i}\int{{\rm d}^4p\over(2\pi)^{4}}
\bigg[
\delta_{\kappa\zeta}
-{\rm i}\int{{\rm d}^4q\over(2\pi)^{4}}
\{G_{\alpha\beta}(q)
G_{\beta\gamma}(q)\}_{\omega}
\Gamma^{\omega}_{\gamma \kappa, \alpha \zeta}(q, p)
\bigg]
\nonumber \\
& & \times
\{G_{\zeta\xi}(p)
G_{\eta\kappa}(p)\}_{\omega}
\lambda^{a}_{\xi\eta}. 
\label{eq:calculation}
\end{eqnarray}
According to the Ward identity (\ref{eq:Ward3}), 
the integrand of the second term 
in (\ref{eq:calculation}) is just 
$\lambda_{\xi\eta} \partial G_{\eta\xi}({\vec p},\varepsilon)
/\partial \varepsilon$ 
so that this term equals zero. 
Substituting (\ref{eq:Ward3}) into the first term 
in (\ref{eq:calculation}), 
we obtain
\begin{eqnarray}
& &\bigg({\partial n \over\partial\epsilon_{a}}\bigg)_{\mu}
= {\rm i}\int{{\rm d}^4p\over(2\pi)^{4}}
\frac{\partial G_{\alpha\beta}^{-1}(p)
}{\partial\varepsilon}
\bigg(\frac{\partial G_{\gamma\delta} ^{-1}(p)
}{\partial\epsilon_{a}}
\bigg)_{\mu} 
\nonumber \\
& &  \quad \quad  \quad
\times
\left[
\{G_{\beta\gamma}(p)
G_{\delta\alpha}(p)\}_{k}
-\{G_{\beta\gamma}(p)
G_{\delta\alpha}(p)\}_{\omega}
\right]. \quad
\label{eq:res}
\end{eqnarray}

Near the Fermi level, (\ref{eq:Gf1}) is expressed in terms of 
quasiparticle as 

\begin{equation}
G^{-1}_{ij}({\vec p},\varepsilon)\approx
a^{-1}_{ij}({\vec p}\,) \varepsilon - 
\epsilon_{ij}({\vec p}\,),
\label{eq:Gf3}
\end{equation}
where 
$a^{-1}_{ij} \equiv a^{-1}_{ij}({\vec p}\,)
=\partial G^{-1}_{ij}({\vec p},\varepsilon)/
\partial\varepsilon|_{\varepsilon=0}$ 
and $\epsilon_{ij} \equiv \epsilon_{ij}({\vec p}\,)$ is given by
\begin{equation}
\left[ \epsilon_{ij}({\vec p}\,) \right] 
=\left[
\begin{array}{cc}
\epsilon_{\rf}-\mu+\Sigma_{\rf\rf}({\vec p},0) & 
V+\Sigma_{\rf\rc}({\vec p},0) \\
V^*+\Sigma_{\rc\rf}({\vec p},0) & 
\xi_{\vec p}+\Sigma_{\rc\rc}({\vec p},0)
\end{array}
\right].
\label{}
\end{equation}
Then the one-particle Green function $G$ can be written in the form
\begin{equation}
G_{ij}({\vec p},\varepsilon) = 
{\tilde A}_{ij}({\vec p}\,)\big/\big[\varepsilon - {\tilde E}_{{\vec p}}^{-}\big]
+\mbox{nonsingular part},
\label{eq:coherent}
\end{equation}
where the renormalization amplitude ${\tilde A}_{ij}({\vec p}\,)$ 
and the dispersion of quasiparticles ${\tilde E}_{\vec p}^{\pm}$ 
near the Fremi level is given by 
\begin{eqnarray}
& &
\big[ {\tilde A}_{ij}({\vec p}\,) \big] = 
{\rm det}\left(\hat{a}\right)
\big/
{\rm tr}\left(\hat{a}\hat{\epsilon}\right)
\nonumber 
\\
& & \quad\quad \times
\left[
\begin{array}{cc}
\xi_{\vec p}+\Sigma_{\rc\rc}({\vec p},0) & 
-V-\Sigma_{\rf\rc}({\vec p},0) \\
-V^*-\Sigma_{\rc\rf}({\vec p},0) & 
\epsilon_{\rf}-\mu+\Sigma_{\rf\rf}({\vec p},0)
\end{array}
\right],
\\
& &
{\tilde E}_{\vec p}^{\pm}\equiv{1\over 2}\left[
{\rm tr}\left(\hat{a}\hat{\epsilon}\right)
\pm\sqrt{\left[{\rm tr}\left(\hat{a}\hat{\epsilon}\right)\right]^{2}
-4 {\rm det}\left(\hat{a}\hat{\epsilon}\right)}\right],
\label{eq:Gf4}
\end{eqnarray}
respectively, with $\hat{a}\equiv [a_{ij}]$ and 
$\hat{\epsilon}\equiv [\epsilon_{ij}]$. 
By (\ref{eq:coherent}), 
the differnce between the $k$-limit and the $\omega$-limit 
of the product $GG$ can be obtained to be
\begin{eqnarray}
& &\{G_{ij}(p)
G_{lm}(p)\}_{k}
-\{G_{ij}(p)
G_{lm}(p)\}_{\omega}
\nonumber
\\
& &= - 2 \pi {\rm i}{\tilde A}_{ij}({\vec k}_{\rm F})
{\tilde A}_{lm}({\vec k}_{\rm F})
\delta({\varepsilon})\delta({\tilde E}_{\vec p}^{-}).
\end{eqnarray}
Substituting this equation into (\ref{eq:res}) 
and taking the summation of the spin variables, we obtain
\begin{equation}
\bigg({\partial n \over\partial\epsilon_{a}}\bigg)_{\mu}
= {\tilde N}_{\rm F}{\tilde A}_{ij}(k_{\rm F})
\bigg(\frac{\partial G_{ji} ^{-1}(k_{\rm F},0)
}{\partial\epsilon_{a}}
\bigg)_{\mu} ,
\label{eq:WardFL}
\end{equation}
where ${\tilde N}_{\rm F}$ is the density of states of the quasi-particles. 
In deriving (\ref{eq:WardFL}), 
we have used the relation ${\rm det}[{\tilde A}_{ij}(k_{\rm F})] = 0$ 
so that 
${\tilde A}_{ij}(k_{\rm F})a^{-1}_{jl}(k_{\rm F})
{\tilde A}_{lm}(k_{\rm F})
={\tilde A}_{im}(k_{\rm F})$. 
By virtue of (\ref{eq:repotential}), 
we finally find that the renormalized potential 
acting on the quasi-particles is given by
\begin{equation}
{\tilde A}^{b}(k_{\rm F}){\tilde u}_{b}({\vec k} \to 0) 
= - \frac{1}{{\tilde N}_{\rm F}}
\bigg({\partial n \over\partial\epsilon_{a}}\bigg)_{\mu}
u_{a}({\vec k} \to 0) ,
\label{eq:potentialFL}
\end{equation}
where ${\tilde A}^{b}(k_{\rm F}) 
= \lambda^{b}_{ij}{\tilde A}_{ji}(k_{\rm F})$. 

Equations (\ref{eq:WardFL}) or (\ref{eq:potentialFL}) 
can be obtaind more easily in the limiting case of heavy electrons, 
$a_{\rf} \equiv a_{11}\ll 1$, $a_{12}=a_{21}^*\sim 0$ and $a_{22} \sim 1$, 
which is the only way for the quasiparticle 
to acquire the extremely heavy mass.  In such a case, the quasiparticles 
consist mainly of f-electrons whose weight in the quasiparticle state 
is given by $1-a_{\rf}|V|^{2}/\xi_{k_{\rm F}}\approx 1$.  Therefore, 
the derivative with respect to $\epsilon_{\rf}$ in (\ref{eq:Ward1}) is 
approximated in a high accuracy, neglecting $a_{\rf}$ compared to 1, 
as 
\begin{eqnarray}
& &-{\partial\over\partial\epsilon_{\rf}}
G_{\rf\rf}^{-1}(p,\varepsilon)\bigg|_{\mu}
\nonumber\\
& &\qquad\approx{1\over a_{\rf}}{\varepsilon-{\tilde E}_{k_{\rm F}}^{+}\over 
\varepsilon-\xi_{k_{\rm F}}}\bigg|_{\varepsilon=0}\times(-1)
{\tilde v}_{k_{\rm F}}^{-}
\left({\partial k_{\rm F}\over\partial\epsilon_{\rf}}\right)_{\mu}
\label{eq:Gf6}
\end{eqnarray}
where the velocity of quasiparticles is defined as 
${\tilde v}_{p}^{-}\equiv\partial{\tilde E}_{p}^{-}/\partial p$, and 
$p$ and $\varepsilon$ have been approximated by $p=k_{\rm F}$ and 
$\varepsilon=0$, respectively.  In deriving (\ref{eq:Gf6}), we have 
considered that the dispersion of quasiparticles near the Fermi level 
is approximated as 
${\tilde E}_{p}^{\pm}\approx {\tilde v}_{p}^{-}(p-k_{\rm F})$ 
and 
\begin{equation}
\lim_{p\to k_{\rm F}}
\left({\partial {\tilde E}_{p}^{-}\over\partial\epsilon_{\rf}}\right)_{\mu}
=-{\tilde v}_{k_{\rm F}}^{-}
\left({\partial k_{\rm F}\over\partial\epsilon_{\rf}}\right)_{\mu}.
\label{eq:Gf6a}
\end{equation}
Since the renormalization factor 
$a_{\rf}$ included in ${\tilde v}_{\rm F}^{-}$ and that in the denominator
of (\ref{eq:Gf6}) cancels with each other, we obtain 
\begin{eqnarray}
-{\partial\over\partial\epsilon_{\rf}}
G_{\rf\rf}^{-1}(p,\varepsilon)\bigg|_{\mu}
&\approx&-v_{k_{\rm F}}
\left({\partial k_{\rm F}\over\partial\epsilon_{\rf}}\right)_{\mu}
\label{eq:Gf7}\\
&\approx&-{1\over N_{\rm F}}
\left({\partial n_{\rf}\over\partial\epsilon_{\rf}}\right)_{\mu},
\label{eq:Gf8}
\end{eqnarray}
where $n_{\rf}$ is the f-electron number density and $N_{\rm F}$ is the 
density of states of non-interacting electrons described by (\ref{eq:PAM}) 
at the Fermi level.  The reason 
(\ref{eq:Gf7}) is approximated by (\ref{eq:Gf8}) is as follows:  
A variation of $k_{\rm F}$ under $\mu$ being fixed corresponds mainly 
to that of f-electron number, because the quasiparticles consist mainly 
of f-electrons\cite{YY} and the change of conduction-electron number is 
limited by the fixed chemical potential $\mu$.  This physical picture 
can be verified on the basis of Gutzwiller approximation applied to PAM 
without $\Ufc$\cite{Rice,Fazekas}.  

By the relations (\ref{eq:Ward2}), (\ref{eq:repotential}), and 
(\ref{eq:Gf8}), the impurity potential acting on f-electrons, 
in the forward scattering limit, is renormalized by the valence 
fluctuations associated with the crossover from Kondo regime to VF one 
as
\begin{equation}
\label{eq:repotential2}
{\tilde u}({\vec k}\to 0)\approx-{1\over N_{\rm F}}
\left({\partial n_{\rf}\over\partial\epsilon_{\rf}}\right)_{\mu}
u({\vec k}\to 0).
\end{equation}
This is an anlog of the relation, (\ref{eq:1}), in which the factor 
$1/z(1+F_{0}^{\rm s})$ is reexpressed as $\chi_{\rm charge}/N_{\rm F}$.  
The relation (\ref{eq:repotential2}) implies that the impurity 
scattering is critically enhanced if the valence of Ce-ion changes 
critically as the f-level $\epsilon_{\rf}$ is tuned, relative to the 
Fermi level, by the pressures.  Indeed, it has been demonstrated 
theoretically that the derivative 
$-(\partial n_{\rm f}/\partial\epsilon_{\rm f})_{n}$ can diverge in the 
system described by the model Hamiltonian (\ref{eq:PAM}) with appropriate 
values of $U_{\rm fc}$ and $\epsilon_{\rm f}$~\cite{Onishi1,Onishi2}.  
This means $-(\partial n_{\rm f}/\partial\epsilon_{\rm f})_{\mu}$ also 
diverges there, because the following relation holds, up to the 
approximation (\ref{eq:Gf8}), 
\begin{equation}
\left({\partial n_{\rm f}\over\partial\epsilon_{\rm f}}\right)_{\mu}
\approx{\displaystyle 
{\left({\partial n_{\rm f}\over\partial\epsilon_{\rm f}}\right)_{n}}\over
\displaystyle
{1-\left({\partial n_{\rm f}\over\partial n}\right)_{\epsilon_{\rm f}}}},
\label{eq:24}
\end{equation}
where the derivative 
$-(\partial n_{\rm f}/\partial n)_{\epsilon_{\rm f}}$ is a small number 
of the order of $z$, the renormalization amplitude.  

In order to see how this enhancement of impurity potential 
affects the behaviors of the resistivity, we need to know the 
$k$-dependence of ${\tilde u}(k)$ for the scattering from 
${\vec p}-{\vec k}/2$ to ${\vec p}+{\vec k}/2$ near the Fermi 
surface.  Although it is not easy to determine the $k$-dependence 
accurately in general, it may be reasonbale to parameterize as 
\begin{equation}
\label{eq:25}
{\tilde u}({\vec k})\approx{1\over \eta+Ak^{2}}
u({\vec k}),
\end{equation}
where $\eta$ is the inverse valence susceptibility, 
$\eta^{-1}\equiv|(\partial n_{\rf}/\partial\epsilon_{\rf})_{\mu}|/N_{\rm F}$, 
and $Ak_{\rm F}^{2}\sim {\cal O}(1)$.  
In the case where the bare impurity potential causes essentially the Born 
scattering, the enhancement of the residual resistivity $\rho_{0}$ 
by the critical fluctuations becomes gigantic.  
Indeed, $\rho_{0}$ is given as 
\begin{equation}
\rho_{0}\approx
\Biggl\langle
{2\pi N_{\rm F}
c_{\rm imp}|u({\vec k})|^{2}(1-\cos \theta)
\over \bigl[\eta+A\bigl(2k_{\rm F}\sin^{2}(\theta/2\bigr)\bigr]^{2}}
\Biggr\rangle_{\rm FS},
\label{eq:26}
\end{equation}
where $c_{\rm imp}$ is a concentration of impurity, 
$\theta$ is an angle between 
${\vec p}\pm{\vec k}/2$, and the on-shell 
condition $\epsilon$=$\xi_{{\vec p}\pm{\vec k}/2}$=0 has been used.  
$\langle\cdots\rangle_{\rm FS}$ means that the average with respect to 
${\vec p}$ over the Fermi surface is taken.  
Here it is noted that explicit dependence of 
renormalization amplitude does not appear due to 
cancellation between that for DOS and that for the damping rate of 
quasiparticles\cite{Langer1}.  Calculation of angular average over $\theta$ 
is performed easily giving rise to 
\begin{equation}
\rho_{0}\propto\ln{1\over \eta}.
\label{eq:27}
\end{equation}
This result remains valid even if we take into account higher order 
terms by calculating the $t$-matrix.  It is because the scattering 
by the renormalized potential (\ref{eq:repotential2}) is 
nothing but the Rutherford scattering in the limit of $\eta\to 0$.  

It should be noted that the present result is not contradictory to 
the Friedel sum rule according to which the scattering probability 
does not diverge so long as the extra charge accumulated locally 
around the impurity is finite\cite{Friedel,Langer2}.  
The form of renormalized impurity potential 
(\ref{eq:repotential2}) becomes long ranged as $\eta\to 0$ 
even though the bare potential is short ranged.  Namely, the change of 
valence near the impurity site extends in long range proportionally 
to $1/r$, $r$ being the distance from the impurity site.  As a result, 
total amount of valence change around the impurity from that of the 
host metal is divergent while the local charge of f- and conduction 
electrons remains finite.  So, the effect of impurity becomes long ranged 
making the scattering probability divergent.  

The expression of $\rho_{0}$ given by (\ref{eq:27}) can explain a huge 
enhancement observed in CeCu$_2$Ge$_2$ and CeCu$_2$Si$_2$ at around 
the critical pressure where the rapid valence change seems to occur.  
This huge enhancement should be compared to the moderate enhancement 
around the magnetic quantum critical point where the enhancement 
arises only through the renormalization amplitude $z$ as discussed in 
Ref.\cite{MN}.  The present mechanism of huge enhancement of $\rho_{0}$ 
is related to other systems near the quantum critical point associated with 
charge instability.  
For example, such a huge enhancement has been observed in Cd$_2$Re$_2$O$_7$ 
at around the pressure where the charge ordering temperature vanishies and 
the superconducting transition temperature $T_{\rm c}$ is considerably 
enhanced compared 
to that at the ambient pressure\cite{hiroi}.  Similar enhancement is expected 
in $\beta$NaV$_6$O$_{15}$ which exhibits pressure induced superconductivity 
of $T_{\rm c}=10$ at around the pressure where the charge ordering 
is suppressed\cite{yamauchi}.  

The enhancement of the two-dimensional charge sucseptibility was observed 
by photoemission spectroscopy near the metal-insulator transition in 
high-$T_{\rm c}$ cuprates LSCO\cite{fujimori}.  This opens a new 
possibility that the doped ions, Sr ions in LSCO, which have only 
subtle influence on electrons in CuO$_2$ plane in the over and optimum 
regions, are transformed into the strong scattering center in the under 
doped region.  Therefore, the carrier doping can suppress $T_{\rm c}$ 
of d-wave superconductivity considerably near the metal-insulator 
phase boundary\cite{MM}.  

\section*{Acknowledgements}
We acknowledge C. M. Varma for stmulationg discussions and S. Fujimoto 
for his comment leading to clarification of implication of the 
present results.  We also acknowledge J. Flouquet and his colleagues 
for encouragements and a hospitality at CEA-Grenoble 
where the last stage of this work was performed.  
This work is supported by the Grant-in-Aid for COE Research (10CE2004) 
from Monbu-Kagaku-sho. 
%the Japanese Ministry of Education, Science, Sports, 
%Culture, and Technology.  


\vskip36pt
\begin{thebibliography}{99}
\bibitem{Kotliar}
G. Kotliar, E. Abrahams, A. E. Ruckenstein, C. M. Varma, 
P. B. Littlewood and S. Schmitt-Rink: 
Europhys. Lett. {\bf 15} (1991) 655.
\bibitem{varma}
C. M. Varma: Phys. Rev. Lett. {\bf 97} (1997) 1535.
\bibitem{MNO}
K. Miyake, O. Narikiyo and Y. Onishi: Physica B {\bf 259-261} (1999) 676. 
\bibitem{MN}
K. Miyake and O. Narikiyo: preprint, cond-mat/0110174, submitted to J. Phys. 
Soc. Jpn.  
\bibitem{Jaccard}
D. Jaccard, E. Vargoz, K. Alami-Yadri and H. Wilhelm: 
Rev. High Pressure Sci. Technol. {\bf 7} (1998) 412; 
D. Jaccard, H. Wilhelm, K. Alami-Yadri and E. Vargoz: 
Physica B {\bf 259-261} (1999) 1.
\bibitem{Thessieu}
C. Thessieu, J. Flouquet, G. Lapertot, A. N. Stepanov and D. Jaccard: 
Solid State Commun. {\bf 95} (1995) 707. 
\bibitem{MMV}
H. Maebashi, K. Miyake and C. M. Varma: preprint, cond-mat/0109276, 
submitted to Phys. Rev. Lett.  
\bibitem{Fulde}
P. Fulde and A. Luther: Phys. Rev. {\bf 170} (1968) 570. 
\bibitem{rosch}
A. Rosch: Phys. Rev. Lett. {\bf 82} (1999) 4280, and references therein.  
\bibitem{Matibet}
O. Betbeder-Matibet and P. Nozi\`eres: Ann. Phys. {\bf 37} 
(1966) 17.  
\bibitem{onodera}
A. Onodera: private communications.  
\bibitem{Muramatsu}
T. Muramatsu et al: J. Phys. Soc. Jpn. {\bf 70} (2001) No.12, in press.
\bibitem{MM}
K. Miyake and H. Maebashi: J. Phys. Chem. Solids {\bf 62} (2001) 53.
\bibitem{Onishi1}
Y. Onishi and K. Miyake: Physica B {\bf 281\&282} (2000) 191.
\bibitem{Onishi2}
Y. Onishi and K. Miyake: J. Phys. Soc. Jpn. {\bf 69} (2000) 3955.
\bibitem{AGD}
A. A. Abrikosov, L. P. Gorkov and I. Ye. Dzyaloshinskii: 
{\it Quantum Field Theoretical Methods in Satistical Physics}, 
2nd editions (Pergamon Press, Oxford, 1965) \S19.3. 
\bibitem{YY}
K. Yamada and K. Yosida: Prog. Theor. Phys. {\bf 76} (1986) 621. 
\bibitem{Rice}
T. M. Rice and K. Ueda: Phys. Rev. Lett. {\bf 55} (1985) 995; 
Phys. Rev. B {\bf 34} (1986) 6420.
\bibitem{Fazekas}
P. Fazekas and B. H. Brandow: Phys. Scr. {\bf 36} (1987) 809.
\bibitem{Langer1}
J. S. Langer: Phys. Rev. {\bf 124} (1961) 1003.
\bibitem{Friedel}
J. Friedel: Phil. Mag. {\bf 43} (1952) 153; 
Nuovo Cimento Suppl. {\bf 7}  (1958) 287.
\bibitem{Langer2}
J. S. Langer and V. Ambegaokar: Phys. Rev. {\bf 121} (1961) 1090.  
%\bibitem{hanazawa}
%M. Hanazawa {\it et al.}/: Phys. Rev. Lett., in press; 
%\bibitem{sakai}
%H. Sakai {\it et al.}/: preprint, submitted to J. Phys. Condens. Matter.
\bibitem{hiroi}
 Z. Hiroi: private communications.  
\bibitem{yamauchi}
T. Yamauchi: private communications. 
\bibitem{fujimori}
A. Ino, T. Mizokawa, A. Fujimori, K. Tamasaku, H. Eisaki, S. Uchida, 
T. Kimura, T. Sasagawa and K. Kishio: Phys. Rev. Lett. {\bf 79} (1997) 2101. 
\end{thebibliography}
\end{document}